\documentclass[12pt]{article}

\usepackage{amssymb}

\newcommand{\remove}[1]{}

\usepackage{subfigure}
\usepackage{graphicx}

\begin{document}




\title{A Flexible Framework for Accurate Simulation \\ of Cloud In-Memory Data Stores}


\author{P. Di Sanzo, F. Quaglia, B. Ciciani, A. Pellegrini\\DIAG - Sapienza Universita' di Roma\\\\D. Didona, P. Romano\\INESC-ID, Lisbon\\\\R. Palmieri, S. Peluso\\Virginia Tech}

\date{}
\maketitle

\begin{abstract}
In-memory (transactional) data stores, also referred to as data grids, are recognized as a first-class
data management technology for cloud platforms, thanks to their ability to match the elasticity requirements imposed by the
 pay-as-you-go cost model. 
 On the other hand, defining the
 well-suited amount of cache servers to be deployed, and the
 degree of in-memory replication of slices of data, in order to optimize
 reliability/availability and performance tradeoffs, is far from
 being a trivial task. Yet, it is an essential aspect of the provisioning process of cloud platforms, given that it
 has an impact on how well cloud resources are actually exploited.
  To cope with the issue of determining optimized configurations of cloud in-memory  data stores, in this article we present a flexible simulation framework offering skeleton simulation models that can be easily  specialized in order to capture the dynamics of diverse data grid systems, such as those related to the specific (distributed) protocol used to provide data consistency and/or transactional guarantees. Besides its flexibility, another peculiar aspect of the framework lies in that it integrates simulation and machine-learning (black-box) techniques, the latter being essentially used to capture the dynamics of the
  data-exchange layer (e.g. the message passing layer) across the cache servers. This is a relevant aspect when considering that the actual data-transport/networking infrastructure 
  on top of which the data grid is deployed
 might be unknown, hence being  not feasible to be modeled via white-box (namely purely simulative) approaches.
 We also provide an extended experimental study aimed at validating instances of simulation models supported by our framework against
 execution dynamics of real data grid systems deployed on top of either private or public cloud infrastructures. Particularly,   our validation test-bed has been based on 
  an industrial-grade open-source data grid, namely Infinispan by JBoss/Red-Hat, and a de-facto standard benchmark for NoSQL platforms, namely YCSB by Yahoo.
 %
 The validation study has been conducted by relying on both public and private cloud systems, scaling the underlying infrastructure up to 100 (resp. 140) Virtual Machines for the public (resp. private) cloud case. 
 \end{abstract}





\section{Introduction}
\label{introduzione}
The advent of
cloud computing has led to the proliferation of a new
 generation of in-memory, transactional data platforms, often referred to as NoSQL data grids, among
 which we can find products such as Red Hat's Infinispan \cite{infinispan-cache-mode}, VMware vFabric GemFire \cite{gemfire}, Oracle Coherence \cite{coherence} and Apache Cassandra \cite{cassandra}.
These platforms well
meet the elasticity requirements imposed by the pay-as-you-go cost model since they
(a) rely on a simplified key-value data model (as opposed to the traditional relational model),
(b)   employ efficient in-memory replication mechanisms
to achieve data durability (as opposed to disk-based logging) and
(c) natively offer facilities for
dynamically resizing the amount of hosts within the platform. 
 They are therefore widely recognized as a core technology for, e.g., emerging big data applications to be hosted in the cloud.


\remove{
Overall,
they allow (non-expert) users to provision a cluster of virtually any size within
minutes.
}

However, beyond the simplicity in their deploy and use,
one aspect that still represents a core issue to cope with when adopting in-memory NoSQL data grids is related to
the (dynamic) resize and configuration of the system. This is of paramount importance in the cloud anytime 
%
%
some predetermined Service Level Agreement (SLA) needs to be matched while also minimizing operating costs related to, e.g., renting the underlying virtualized infrastructure. However, accomplishing this goal is far from being trivial, as forecasting
the scalability trends of real-life, complex applications deployed on distributed in-memory transactional platforms is
 very challenging.
 In fact, as also shown in \cite{taas},
when the number of nodes in the system grows and/or the workload intensity/profile changes, the performance of these platforms may exhibit strong non-linear behaviors, which are imputable to the simultaneous, and often inter-dependent, effects of contention affecting both physical (CPU, memory, network) and logical (conflicting data accesses by concurrent transactions) resources.

Recent approaches have tackled the issue of predicting the performance of these in-memory data grid platforms (e.g. to assist dynamic reconfiguration processes) by relying on
analytical modeling, machine learning or a combination of the two approaches (see, e.g.,
\cite{180172,taas}). In this article we provide an orthogonal solution which is based on the combination of discrete event simulation and machine learning techniques.

Specifically, we provide a framework for instantiating discrete event models of data grid platforms, which can be exploited for what-if analysis in order to determine what would be the effects of reconfiguring various parameters, like: (i) the number of cache servers within the platform;
(ii) the degree of replication of the data-objects;
(iii) the placement of data-copies across the platform.
Hence, it can be used in order to determine well suited configurations (e.g. minimizing the cost for the underlying virtualized infrastructure) vs variations of the volume of client requests, the actual data conflict and the locality of data accesses. It can also be used for long term SLA-driven planning in order to determine whether 
the data grid can sustain an increase in the load volume and at what operational cost - as a reflection of the increased amount of resources that shall be provisioned from the cloud infrastructures.

%

The framework has been developed as a C static library
implementing data grid models developed according to the
traditional event-driven simulative approach, where the evolution of each individual entity to be simulated within the model is expressed by a specific event-handler ({\footnote{The actual code implementing the framework is freely available for download at the URL http://www.dis.uniroma1.it/{\textasciitilde}hpdcs/software/dags-with-cubist.tar}). On the other hand, the library has been structured in order to allow easy development of models of data grid systems offering specific facilities and supporting specific data management algorithms (e.g. for ensuring consistency of replicated data). As for this aspect, distributed data grids relying on two-phase-commit (2PC) as the
native scheme for cache server coordination, as typical of most of the mainstream implementations (see, e.g., \cite{infinispan-cache-mode}), have an execution pattern already captured by the skeleton model offered by the library. Hence, models of differentiated
2PC-based data management protocols could be easily implemented on top of the framework. Further, models natively offered within the framework include those of data grids ensuring repeatable read semantics, which are based on lazy locking. Models of primary data ownership vs multi-master schemes are also natively supported.

The ability of our simulation framework to reliably capture the dynamics of data grid systems deployed in real cloud environments is strengthened by the combination of the white-box simulative approach with black-box machine learning techniques. The latter 
aim to capture (and to predict) the data-transport/networking sub-system dynamics.
This kind of integration spares us (and any framework user) from
the burden of  explicitly modeling the dynamics of the network layer within the simulation code, which is known to be
an error-prone task given the complexity and heterogeneity of
existing network architectures and/or message-passing/group-communication systems \cite{CouceiroRR10} (\footnote{Group communication systems such as \cite{jgroups} are often used as data exchange layers within real data grid products. They typically exhibit complex dynamics that can vary on the basis of several parameters, hence being difficult to be reliably captured via white-box models.}). Also, the reliance on machine learning
for modeling network dynamics widens the framework practical usability in modeling 
%
data grid systems deployed over  virtualized
cloud environments where users have little or no knowledge of the underlying network
topology/infrastructure and of how the lower level message passing sub-systems are structured. For these scenarios, 
the construction of white-box simulative models would not only be a complex task, rather it would be 
unfeasible.


Fidelity of the framework in modeling the dynamics of real systems is 
demonstrated via a case study where we compare simulation outputs with  measurements obtained 
running the YCSB benchmark by Yahoo \cite{ycsb}, in different configurations, on top of the Infinispan data grid system by JBoss/Red-Hat \cite{infinispan-cache-mode}, namely the mainstream data layer for the JBoss application server. 
We note that the YCSB benchmark has been designed to explicitly assess the run-time behavior of cloud data stores, and has been already exploited as a reference in a set of recent studies (see, e.g., \cite{taas}), hence looking as an ideal candidate for the our validation study.
 Also, Infinispan supports distributed data management schemes that can be considered as instances of ``archetypal'' ones, which strengths the relevance of our study in assessing the actual quality of the models that can be instantiated via the framework. Further, the experiments have been conducted by relying on both private and public (namely FutureGrid \cite{futuregrid}) cloud systems, by scaling the underlying infrastructure up to 140 Virtual Machines for the private cloud, and up to 100 Virtual Machines for the public one.
By the validation study, the framework provides (at least) 80\% accuracy in predicting core performance metrics such as the system throughput across all the tested configurations, and on the order of 95\% accuracy for most of them. 


The remainder of this paper is structured as follows. In Section \ref{related} we discuss related work.
The framework organization is presented in Section \ref{framework}. Experimental data are reported in
Section \ref{data}.
}
\section{Related Work}
\label{related}

The issue of optimizing the configuration of data grids
has been addressed in literature according to differentiated methodologies. The recent works in \cite{taas,DidonaFHRS13,mascots2014} provide approaches where analytical modeling and machine learning are jointly exploited in the context of performance prediction of data grid systems hosted on top of cloud-based infrastructures. The analytic part is mainly focused to capturing dynamics related to the specific concurrency control algorithm adopted by the data grid system, while machine learning is targeted at capturing contention effects on infrastructure-level resources. Differently from our approach, these works cope with specific data grid configurations (e.g. specific data management algorithms and/or specific workload profiles) to which the analytical models are
targeted. For example, they assume arrivals of transactions to the system to form a Poisson process; however, recent works suggest that, in large scale data centers, the inter-arrival time of requests to a data grid may not follow the exponential distribution~\cite{Atikoglu:2012}. In the same guise, those models are bound to specific data access pattern dynamics (e.g., in terms of data locality), which are not general enough to encompass complex data-partitioning schemes across the servers~\cite{CurinoZJM10}.
Instead, we offer a framework allowing the user to flexibly model, e.g., differentiated data management schemes without imposing specific assumptions on the workload and data access profile (in fact real execution traces can be used to drive the
simulated data access).


The proposals in \cite{ncca2012,ncca2014} are based on the exclusive usage of machine learning, hence they provide performance prediction tools that do not have the capability to support what-if analysis in the wide (e.g. by studying the effects of --significant-- workload shifts outside the workload-domain used during the machine learning training phase).
Rather, once a machine learning-based model is instantiated via these tools, it stays bound to a specific scenario (e.g. to a specific deploy onto a given infrastructure), and can only be used to (dynamically) reconfigure the target data grid that has been modeled. We retain similar capabilities; however, by limiting the usage of the machine learning component to predicting messagging/networking dynamics, we also offer the possibility to perform what-if analysis and exploration of non-instantiated configurations (e.g. in terms of both system setting and workload profile/intensity).


One approach close to our proposal has been presented in \cite{toolkit}. This work presents a simulation layer entailing the capabilities of simulating data grid systems. Differently from this proposal, which is purely simulative, our approach exhibits higher flexibility in terms of its ability to reliably model the dynamics of data grid systems in the cloud thanks to the combination of simulative and machine learning approaches. In fact, as already pointed out, the machine learning part allows for employing the framework in scenarios where no (detailed) knowledge on the structure/internals of the networking/messaging system to be modeled is provided to the user. As for this aspect, the usage of machine learning for the performance prediction of group communication systems has been pioneered in \cite{CouceiroRR10}. However, the idea of combining simulative and machine learning-based models is, to the best of our knowledge, still unexplored in the literature.

Simulation of data grid systems has also been addressed in \cite{simutools-best-2011}. In this proposal,
the modeling scheme of the data grid is based on Petri nets, which are then solved via simulation. 
With respect to this solution, we propose a functional model that does not explicitly rely on
modeling formalisms, except for the case of the CPU, which is modeled via queuing approaches rather than Petri nets. Further, one relevant difference between the work in \cite{simutools-best-2011} and our proposal lies in that our simulation models are able to simulate complex transactional interactions entailing multiple read/write (namely get/put) operations within a same transaction. Instead, the work in \cite{simutools-best-2011} only models single get/put interactions to be issued by the clients, thus making 
 our approach more general.

Also related to our proposal are the simulation models developed in \cite{CalheirosRBRB11}. However,  unlike this article, the focus of that work is on modelling lower levels dynamics related to IaaS management (e.g., scheduling of VMs to a set of physical resources).
Finally, a work still marginally related to our proposal can be found in \cite{simutools-2012-dalle}, where a simulation environment for backup data storage systems in peer-to-peer networks is presented. Compared to our proposal, this work is focused on lower level data management aspects, such as the explicit modeling of actual stable storage devices. Instead, our focus is on distributed dynamics at the level of in-memory data storing systems, which are essentially independent of (and orthogonal to) those typical of stable storage technologies.

\section{The Framework}
\label{framework}

\begin{figure}[t]
\centering
 \includegraphics[width=10cm]{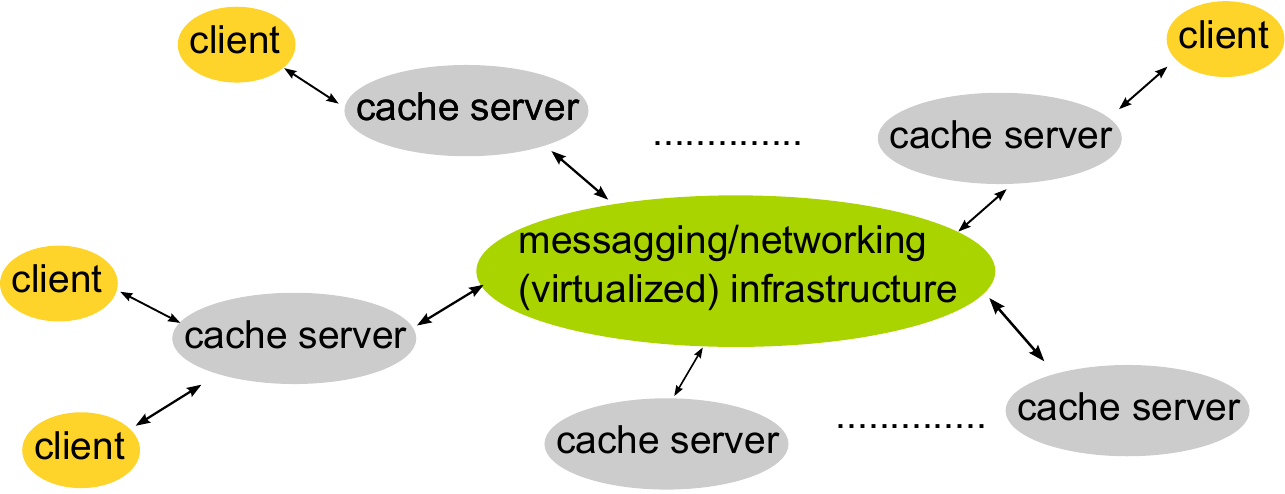}
\caption{Reference system architecture.}
 \label{architecture}
\end{figure}

The data grid architectures we target in
our framework can be schematized (at
high level) as shown in Figure \ref{architecture}. In particular, they are essentially composed of two types of entities, namely:

\begin{itemize}
\item {\em cache servers}, which are in charge of maintaining copies of entire, or partial, data-sets;
    \item {\em clients}, which issue transactional data accesses and/or updates towards the cache servers.
\end{itemize}


The cache servers can be configured to run different distributed protocols in order to guarantee specific levels of isolation and data consistency while supporting transactional data accesses. For instance, the 2PC protocol can be exploited in order to guarantee atomicity while updating distributed replicas of the same data-object, as it typically occurs in commercial in-memory data platform implementations (see, e.g., \cite{infinispan-cache-mode}).
 Also, an individual transactional interaction issued by any client can be mapped onto either a single
 {\sf put}/{\sf {get}} operation of a data-object, or a more complex transactional manipulation involving several {\sf put}/{\sf {get}} operations on multiple data-objects, which is demarcated via {\sf begin} and {\sf end} statements.

As already mentioned, our framework has been designed in order to be layered on top of a combination of simulative models and machine learning ones.  The simulative part of our data grid models is mapped onto a system representation only entailing two types of simulation objects: (A) client objects and (B) cache server objects. On the other hand,
network dynamics and the associated transfer delays for messages sent across the different components within the simulation model are not simulated by explicitly including some network simulation-object. Instead, a kind of machine learning oracle is queried while the simulation is in progress in order to determine the expected latency for message delivery, depending on parameters such as the system scale, the message size and the current system load. In fact, higher volumes of concurrent data accesses may lead to scenarios where the messaging layer at the bottom of the software stack characterizing the data grid system would tend to be stressed more than what would happen with lower load volumes. This is because more coordination actions across the cache servers are requested per time unit due to  either (i) data transfer (e.g., in case the local cache server 
does not already store locally  a replica of the data slice accessed in read mode by a locally handled transaction),
 or (ii) the handling of the commit phase of the transaction (since more control messages needs to be concurrently handled by the messaging system during, e.g., the 2PC-based commit phase). Also, the message size, which may in turn impact the delivery delay of the messaging system, depends on the amount of per-transaction accessed data. In fact it is typical that data grid systems handle the transaction commit phase by transferring information on (at least) the write set of the transaction across the involved cache servers.

\remove{
 As for the latter point, given that our base simulated protocol for cache server coordination is 2PC, upon any 2PC-based coordination action, only those cache servers maintaining copies of the data to be locked need to mutually exchange
simulation events. If the degree of replication is limited, as typical of partial replication schemes, the set of servers that coordinate with each other while handling
client requests gets reduced, hence leading to simulate the evolution of groups of cache servers along different critical paths in different phases of the simulation run. Again, this likely favors parallelism in the model execution.
}

In the next subsections we initially focus on the structure and discrete-event patterns of cache server and client simulation objects. In particular, we focus on 
the corresponding skeleton exposed by the framework and on the support for easy modifiability of the simulated logic so as to allow easy (re-)implementation of differentiated data grid simulation models, particularly w.r.t. different concurrency control schemes and protocols for distributed transaction atomicity. Successively, we enter the details of the machine learning approach used to model message delivery latencies across the system components, and of its integration with the simulative part of the framework.

\begin{figure}
\centering
\includegraphics[width=.750\textwidth]{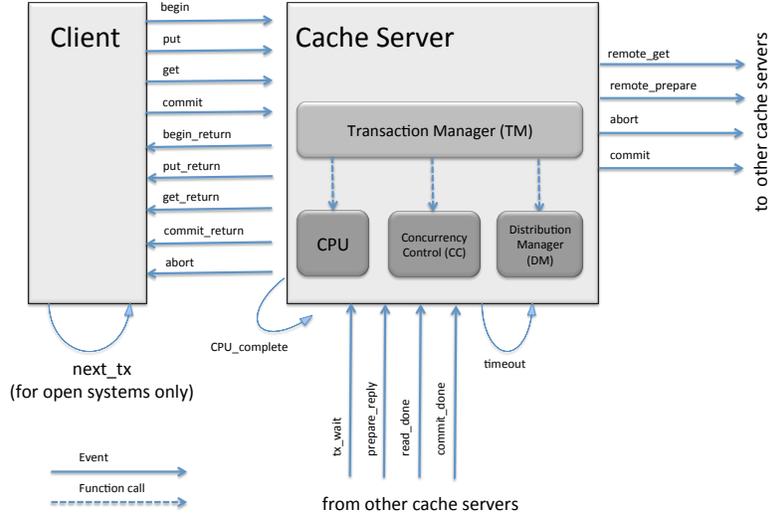}
\vspace*{-0.3cm}
    \caption{Client and Cache-Server Simulation Objects.}
\label{cache-server}
\end{figure}

\subsection{The Cache Server Simulation-Object}

A cache-sever simulation object can be schematized as shown in Figure \ref{cache-server}.
By the scheme we can identify four main software components:
\begin{itemize}
\item the transaction manager (TM);
\item the distribution manager (DM);
\item the concurrency control (CC); and
\item the CPU.
\end{itemize}

Any simulation event destined to the cache server is eventually passed as input to TM, which acts therefore as a front-end for event processing. Upon the scheduling of any event, TM determines the amount of time required to process the requested activity, which depends on the type of the scheduled event, and on the current CPU load.
Then, the CPU load is updated on the basis of the newly scheduled activity.
Additionally, the completion time
for the activity is determined, which depends on the current CPU load. Finally, a {\sf CPU-complete} event is scheduled at the corresponding simulation time. 

To determine the CPU processing delay, the CPU has been modeled as a G/M/K queue, which allows capturing scenarios entailing multiple CPU-cores. Although more sophisticated  models could be employed (see, e.g., \cite{Fujimoto87}), we relied on
G/M/K queues since, in our target simulation scenarios, the core dynamics of interest are the ones related to contention on logical resources, namely data-objects, rather than physical resources, and to distributed (locking) strategies for the management of atomicity of the updates of distributed/replicated data copies. Hence, distributed coordination delays play a major role in the determination of the achievable performance, as compared to CPU delays for processing local activities. As a consequence, the G/M/K queue is expected to be a fairly adequate model for the objectives of the framework.
For the same reason depicted above, effects by virtual memory on the latency of operations provided within the data grid simulation model are not explicitly considered.
\remove{
\begin{figure}
\begin{scriptsize}
\begin{verbatim}
to be done
\end{verbatim}
\end{scriptsize}
\caption{Architecture of the Cache-Server Simulation Object.}
\label{cache-server}
\end{figure}
}

When a local processing activity is completed, TM takes again control (via the aforementioned {\sf CPU-complete} event) and performs the actual updates related to the activity. These updates are different depending on the exact type of event that triggered CPU work.

 As for events scheduled by client simulation objects towards the cache servers, the corresponding event-types
 within the framework skeleton are listed below:
\begin{itemize}
\item {\sf begin}, used to notify TM that a new transactional interaction has been issued by some client, which must be processed by the cache server;
\item {\sf get}, used to notify that a read operation on some data object has been issued by the client within a transaction;
\item {\sf put}, used to notify that a write operation on some data object has been issued by the client within a transaction; and
\item {\sf commit}, used to indicate that the client ended issuing operations within a transaction, whose commit can therefore be attempted.
\end{itemize}

The handling of the {\sf begin} event at the side of TM 
is implemented via the internal function {\tt setupTransaction}, which simply takes as input
the current simulation time and
pointers to two records of type {\tt TxInfo} and {\tt TxStatistics}. These records are both automatically allocated by the framework and linked to the corresponding records of already active transactions.

The actual internal structure of both the {\tt TxInfo} and {\tt TxStatistics} records can be defined by the simulation modeler. In fact, the framework provides a proper header file, named {\tt transaction.h}, where the modeler can specify such structure.
The only constraint is that the top standing field of
{\tt TxInfo}, must be of type {\tt TxId}, which keeps the transaction unique identifier, automatically generated by the cache server
just to facilitate the actual management within model execution.

This is one of the core facilities on top of which lies the framework flexibility wrt the actual model implementation. In fact, with this organization, the modeler can keep track of management information (i.e. {\tt TxInfo}) and statistics information (i.e. {\tt TxStatistics}) associated with active transactions within whichever modeler-defined data structure, which is automatically allocated and managed by the framework into the heap.

The reason for allowing the modeler to exploit two different data types lies in that the content of {\tt TxInfo} is made valid according to a cross cache-server scheme. In fact, it is automatically transferred
to remote cache server simulation objects when cross scheduling of events is actuated, as we shall discuss. This is relevant in any simulated scenario where some transaction set-up information needs to be made available to remote cache servers, e.g., for distributed contention management purposes.

On the other hand, the content of {\tt TxStatistics} is not transferred across different simulation objects, being it locally handled by the cache server acting as the coordinator of the transaction. In particular, upon finalization of a transaction, TM automatically invokes the module {\tt finalizeTransaction}, which receives as input the current simulation time, and again pointers to both {\tt TxInfo} and {\tt TxStatistics} records so to allow for their update (particularly the statistics).
The release of these buffers within the framework is again handled automatically. However, before releasing any of them, a special module {\tt statisticsLog} is called, passing as input pointers to both of them, allowing the modeler to finally log, e.g. onto the file system, any provided statistical data.


\remove{
 This data structure can keep track of both audit data for the generation of statistics (such as transaction startup and completion time) as well as information for contention management to be exploited at the level of the CC module. In our already developed data grid models, we have exploited this framework facility in order to define {\tt TxInfo} such in a way to log per-transaction audit information.
}

As for {\sf get} and {\sf put} simulation events, they cause the TM module to simply query (via synchronous procedure invocation) the DM module. This is done in order to get information about what cache servers figure as the owners of the data object to be accessed. In our architecture, the DM module provides this information back in the form of a pointer to a list of cache server identifiers (hence simulation object identifiers), where each record also keeps additional information specifying whether a given cache sever is (or is not) the primary owner of a copy of the data object to be accessed. Once TM gets this information, it then determines the pattern of additional simulation events to be scheduled. More in details, primary ownership has relevance for {\sf put} (namely write) operations on data objects. Instead, {\sf get} operations are not affected by the presence of a primary owner, if any. Let us discuss this aspect in detail.

In case of {\sf get} simulated events, the cache sever determines whether it is the owner of a copy of the data object. In the positive case, the read operation on the data object will simply result in an invocation of the CC module on this same cache server instance. Otherwise, {\sf remote\_get} simulation events are scheduled for all the cache servers figuring as owners of a copy of the data object. As we shall see in Section \ref{machine-learning}, {\sf remote\_get} events are scheduler at later time to model the corresponding  request transmission delay as determined by a query to the machine-learning component included in the framework.
 Upon their execution at the destination cache severs, which will still entail passing through the simulated CPU processing stage, these events will trigger CC invocations on those cache server simulation objects.

One important aspect associated with the above scheme is that the {\sf get} operation may be blocked at the level of CC, depending on the actual policy for controlling concurrency. On the other hand, even in case of CC simulated algorithms implementing non-blocking read access to data (as is the case for most data grid products guaranteeing weak data consistency, such as read committed or repeatable read semantics \cite{infinispan-cache-mode}),
the read operation may anyway be blocked in case no local copy exists and needs to be fetched by some remote cache sever. This is automatically handled by our framework since the TM module
records information on any pending simulated read operation within a proper data structure. When setting up the record for a given operation, information on the remotely-contacted cache servers, if any, is also installed. That record will be removed only after processing the corresponding reply simulation events from all those cache servers, which is done for allowing an optimized execution flow for those reply events. On the other hand, the operation is unlocked (and a reply event is scheduled towards the corresponding client) when the first copy of the data becomes available from whichever cache server, hence after processing the first simulation event associated with a read-reply. Note that this architectural organization automatically covers the case where the transaction operation is blocked locally, due to the current state of CC. In such a case, the contacted-server list will be filled with the identifier of the local cache server, and a read-reply event from this same server (which will be scheduled by the local CC module, as we shall discuss) will be used to unlock the request and to schedule the reply towards the client simulation object.

\remove{
A particular case is when $n$ is set to 1, which means that a reply to the client becomes available as soon as the fastest cache server replies. Anyway, $n$ is a simulation modeler defined parameter, whose choice can be coupled with the specifically nature of the CC algorithm.
Even though the actual data grid models we have already implemented and tested refer to non-blocking CC treatment for read operations with $n$ set to 1, the above structuring of the framework provides high flexibility, and the possibility ti define different un-blocking scenarios for .
}

In case of {\sf put} operations (namely data object updates) the corresponding simulation events only trigger the update of some meta-data locally hosted by the cache server, which are embedded into records treated at the same manner as the above-mentioned (modeler-defined) {\tt TxInfo} record. These include the operation identifier, and the key associated with the data object to be updated. This behavior simulates a simple local update of the transaction write set, which is again reflected into a cross cache sever valid record (in case cross server events for that transaction are scheduled) which we name {\tt TxWriteSet}. 

On the other hand, the meta-data are queried upon simulating a {\sf get} operation to determine whether the data object to be read already belongs to the transaction read/write set (hence whether the get operation can be served immediately via information within the read/write set). In such a case, the simulation-event pattern for handling the {\sf get} is different from the general one depicted above since it only entails simulating local CPU usage required for providing the value extracted from the transaction read/write set to the client.
This implicitly leads the framework to provide support for simulating transactional data management protocols ensuring at least repeatable-read semantic.

More complex treatments are actuated when handling {\sf commit} simulation events incoming at the cache servers. In particular, differentiated simulation event patterns are triggered by TM depending on whether the simulated scheme entails a primary owner for each data object or not. For primary ownership scenarios, the {\sf prepare} will result in scheduling {\sf remote\_prepare} events towards all the primary cache servers that keep copies of the data to be updated (each event carries the keys associated with the data objects to be updated, which are again retrieved via the {\tt TxWriteSet} data structure maintained by the cache server acting as transaction coordinator). TM can determine this set of cache servers by exploiting the keys associated with the written data objects (which are kept within the transaction write set). If one of these cache servers corresponds to the server currently processing the prepare request, then, after passing through the CPU processing stage, the local CC module is immediately invoked. At this point we are in a situation similar to the one depicted above for the case of read access to remote data. In particular, for the preparing transaction, the framework
logs the identities of the contacted servers, and then waits for the occurrence of {\sf prepare\_reply} simulation events scheduled by any of these servers. For homogeneity, even when one of the contacted CC module is the local one, the reply from this module occurs via the scheduling of such a {\tt prepare\_reply} event, thus giving rise to the situation where the CC module exhibits the same simulated behavior (in terms of notification of its decisions) independently of whether the prepare phase for the transaction
needs to run local tasks on the same cache server, or remote tasks. Hence, the CC module operates seamless of any simulated data distribution/replication scheme. The above simulation-event pattern is only slightly varied in case of non-primary ownership of data objects since the framework
will schedule these {\sf prepare} events for all the servers keeping copies of the data to be updated. This again allows the CC module to operate transparently to the ownership scheme.

For both the schemes, in case the {\sf prepare\_reply} events are positive from all the contacted servers, final {\sf commit} events are scheduled for all of them, which will ultimately
result in invocations of the CC module. On the other hand, {\sf abort} events are scheduled in case of negative prepare outcome. Further, for the case of primary ownership, the {\sf commit} events are propagated to the non-primary owners, in order to let them reflect data update operations.

Let us now detail the behavior of the CC simulation model, which represents one core component of our framework architecture. By the above description, this module is invoked upon the occurrence of {\sf get} or {\sf remote\_get} events, {\sf remote\_prepare} events, and {\sf commit} events. However, all these events are actually intercepted and initially processed by the TM module which, as said, is the front end simulation-handler within the cache server simulation object. Hence, ultimately, the CC module is oblivious of  whether a requested action is associated with some local or remotely-executed transaction. It only takes the following input parameters:

\begin{itemize}
\item
a pointer to the {\tt TxInfo} record (recall that, in the simulation flow, the field {\tt TxId} at the top of this record has been automatically set by TM upon processing the {\sf begin} event, while additional transaction information can be defined by the modeler by setting it via the {\tt setupTransaction} module);
\item a pointer to {\tt TxStatistics} (or NULL if the cache server is not the transaction coordinator);
\item the {\tt type} of the operation to be performed (read, prepare or commit);
\item the {\tt key} of the data object to be involved in the operation (this is for read operations); and
\item the {\tt TxWriteSet} to be used for CC purposes (this is for the prepare case).
\end{itemize}

On the other hand, CC can reply to invocations by generating one or more of the events listed below towards TM:

\begin{itemize}
\item {\tt TX\_WAIT}, indicating that the currently requested operation leads to a temporary block of the transaction execution;
\item {\tt READ\_DONE}, indicating that the data object can be returned to the reading transaction;
\item {\tt PREPARE\_DONE}, indicating that the transaction has been successfully prepared;
\item {\tt PREPARE\_FAIL}, indicating that the transaction prepare stage has not been  completed correctly; and
\item {\tt COMMIT\_DONE}, indicating that the transaction commit request has been processed.
\end{itemize}

Each of the above events is not directly routed towards the destination simulation object (hence these events are not actual simulation events, rather only event generation indications), just because CC is not aware of whether they must represent replies for the local cache server or remote cache servers, or even the client. Hence, within the framework they are intercepted by a dedicated layer, which buffers these CC triggered event-generation requests 
 so as to make them available for actual scheduling (towards the correct destinations). The latter is actuated by the TM module once it takes back control upon the return of CC. As such, the events triggered by CC can be re-mapped onto actual simulation events to be exchanged across different simulation objects. As an example, {\tt PREPARE\_DONE} and {\tt PREPARE\_FAIL} events are re-mapped and actually scheduled as the aforementioned {\tt prepare\_reply} events, with proper payload (indicating positive or negative prepare outcomes). Further, the CC module can raise the request for issuing {\tt TIMEOUT} events, which can be useful in scenarios where CC actions are also triggered on the basis of passage of time.

Overall, the simulation modeler is easily allowed to implement different concurrency control algorithms by completely ignoring data distribution and replication schemes.  She only needs to deal with transaction identifiers, basic transaction setup information and relations across different transactions, on the basis of the actual data objects locally hosted by a given cache server. 
This is a relevant achievement when considering that great research effort is currently being spent in the design of concurrency control algorithms suited for cloud data stores, which provide differentiated consistency vs scalability tradeoffs (see, e.g., \cite{Pelu11b,DBLP:conf/srds/ArdekaniSS13,DBLP:conf/europar/ArdekaniSSP13,DBLP:conf/sosp/SovranPAL11,DBLP:conf/middleware/PelusoRQ12}), each one fitting the needs of different application contexts. Having the possibility to provide simulation models for such differentiated algorithms by exploiting our framework can definitely reduce the time and effort required for assessing their potential.

To determine what are the locally hosted data objects, hence the locally hosted keys, CC accesses a hash table that gets automatically setup upon simulation startup. On the other hand, the meta-data required to keep relations across active transactions, (e.g. wait-for relations), and the corresponding data structure is completely left to programming by the simulation-modeler. It can be again defined, in terms of types, within the {\tt transaction.h} header file. However, the actual instance of this data structure can be accessed via a special pointer which is passed to the CC module by the framework as an additional input parameter. We note that if the pointer value is {\tt NULL}, then CC has not yet allocated and initialized the structure, hence this must be done, and the actual pointer to be used in subsequent calls to CC can be setup and returned upon completion of the current CC execution.

Let us go back to the {\sf TxInfo} record. As we have said, this is modeler-defined and can keep track of per-transaction meta-data, which can be exploited by the CC module in order to support the actual concurrency control logic. 

Let us finally consider two different examples of how to model via the framework different CC algorithms. One is a classical 2PC based data-grid CC algorithm where every transaction is successfully prepared at any site in case the target data object to be updated is not currently locked upon the prepare request. On the other hand, the second scenario shows how to model cases where the transaction is prepared only in case the target data has a timestamp lower than the transaction timestamp. The examples are presented via pseudo-code for simplicity.


\subsubsection{Example One: Base 2PC} In Figure \ref{example-a} we show the pseudo-code defining the entries of {\tt TxInfo} and some part of the core logic at the level of CC. In this case, {\tt TxInfo} is not required to keep
transaction control information targeted  at contention 
management; it simply maintains transaction identification information.
 On the other hand, the base setup  for
concurrency management  can be actuated by simply setting up a wait-for table where transaction identifiers are queued in different rows depending on what other transaction holds the lock they would like to get on a given data object (the top standing transaction is therefore the one to which the lock has been granted).
In the pseudo-code we show a scheme where, upon simulating a prepare request, the associated transaction is always queued. On the other hand, upon commit or abort events for a pending transaction, the subsequent transaction in the wait-for list is reactivated, with positive reply to the original prepare request.

\begin{figure}[t]
\begin{footnotesize}
~\\\hrule
\begin{verbatim}

record TxInfo{
    TxId
    ...
} //end record

CC-logic(input: task T, pointer CC-Table){

if (CC-table == NULL)
    allocate and initialize [wait-for,active-tx] table;
    // keys are data object identifiers or TxId values
    // entries are lists of TxInfo records or TxId values
    set CC-table point to the allocated table

case T.type
    prepare:
       link T.TxInfo.TxId to CC-Table.active-tx
       AllPrepareKeys = T.TxWriteSet
       link T.TxInfo to CC-Table.wait-for[AllPrepareKeys]
       if T.TxInfo not top standing for some key
          generate event TX_WAIT[T.TxInfo]
          generate event TIMEOUT[T.TxInfo]
       else generate event PREPARE_DONE[T.TxInfo]
    ....
    timeout:
    commit:
       unlink T.TxInfo.TxId from CC-Table.active-tx
       unlink T.TxInfo from CC-Table[AllOccurrences]
       if (T.type == commit) generate COMMIT_DONE[T.TxInfo]
       else generate PREPARE_FAIL[T.TxInfo]
       for all TxInfo top standing in CC-Table[AnyPresenceRow]
           generate event PREPARE_DONE[TxInfo]
    ....

return CC-Table
} //end CC
\end{verbatim}
\hrule
\end{footnotesize}
\caption{Example One.}
\label{example-a}
\end{figure}


\subsubsection{Example Two: Timestamp Based 2PC}
In Figure \ref{example-b} we show the pseudo-code defining the entries of {\sf TxInfo} and some parts of the core logic at the level of CC, where this time we have a variation that leads the {\tt TxInfo} record to keep cross-server control information specifically targeted at data contention management, namely a timestamp value. In this case, differently from the previous scenario, a transaction for which a prepare event has been issued can get successfully prepared only in case its timestamp is greater than the timestamp of any data object accessed in write mode. We note that in this scenario, the CC module, upon setting up the {\tt CC-Table}, needs to take care of setting up meta-data for the explicit maintenance of data object timestamp values.

\remove{
Compared to the previous example, this time the

In this case, {\sf TxInfo} is not required to keep
transaction control information, but only audit information. On the other hand, the base setup  for
concurrency management  can be actuated by simply setting up a wait-for table where transaction identifiers are queued in different rows depending on what other transaction holds the lock they would like to get on a given data object (the top standing transaction is therefore the one to which the lock has been granted).
In the pseud-code we show a scheme where, upon the handling of any prepare request, then the associated transaction always gets queued. On the other and upon commit or abort events for a pending transaction, then the subsequent one in the wait-for list gets reactivated, with positive reply to the original prepare request.
}

\begin{figure}[t]
\begin{footnotesize}
~\\\hrule
\begin{verbatim}
record TxInfo{
    TxId
    timestamp
    ...
} //end record

CC-module(input: task T, pointer CC-Table){

if (CC-Table == NULL)
    allocate and initialize [wait-for,DOT] table;
    // DOT stands for data-object-timestamp
    // table access keys are data object identifiers
    // entries are lists of TxInfo records or DOT values
    set CC-Table point to the allocated table

case T.type
    prepare:
        AllPrepareKeys = T.TxWriteSet
        if T.TxInfo.timestamp > CC-Table.DOT[AllPrepareKeys]
           link T.TxInfo to CC-Table[AllPrepareKeys]
        else generate event PREPARE_FAIL[T.TxInfo]
            goto out
        if T.TxInfo not top standing for some key
          generate event TX_WAIT[T.TxInfo]
          generate event TIMEOUT[T.TxInfo]
       else generate event PREPARE_DONE[T.TxInfo]
    ....

out:
 return CC-Table
} //end CC
\end{verbatim}
\hrule
\end{footnotesize}
\caption{Example Two.}
\label{example-b}
\end{figure}

\remove{

, simulation frameworks provides different event patterns and interactions depending on whether we are in presence of a primary owner or not

\paragraph{Handling {\sf get} operations}

Specifically, in case of primary ownership by a cache server, that cache sever will be the target for the corresponding {\sf get} or  {\sf put} operation. In case the primary owner cache server corresponds to the one where the client operation has been issued, then the data object is simply accessed locally, which results in a synchronous invocation to the CC module. As we shall discuss, this may lead the transaction to experience a wait phase depending on the current state at the level of CC. On the other hand, in case the primary owner is a remote cache server, then a {\sf remote\_get} or {\sf remote\_put} simulation event gets scheduled for the corresponding simulation object at a later time (which models the corresponding  message transmission delay). Once executed on the destination cache severs, which will still entail passing through the CPU processing stage, these events will trigger CC invocations on the corresponding  cache server simulation objects. Hence, local or remote read write operations on data objects are ultimately treated homogeneously.

{\sf get} or {\sf put} operation will be resolved in a {\sf local\_get} or {\sf local}

 be specified by the simulation model developer.

}

\subsection{The Client Simulation-Object}

Client simulation objects have an internal structure that does not need to be changed by the simulation modeler. In fact, she only needs to specify, via configuration files within the framework, what type of probability distribution must be used for determining the data to be accessed, and what distributions need to be used for determining the number of operations to be executed within a transaction and the type (read or write) of each operation.

As for this aspect, the framework already offers the possibility to use differentiated access distributions, some of which are analytic, while others have been determined by relying on traces of known benchmarks.
Further, the clients can be configured in order to simulate either an open or a closed system. For the former case, the simulation modeler needs to specify the rate of generation of transactions at the client side.
As a final note, our clients also embed the possibility to generate the workload by directly relying on traces (rather than on distributions derived from the traces).

\subsection{Modeling Message Exchange Dynamics via Machine-Learning}
\label{machine-learning}

As hinted, our framework relies on black-box, machine-learning-based modeling techniques to forecast the
dynamics at the level of the message-passing/networking  sub-system.
 Developing
white-box models (e.g. simulative models) capable of capturing accurately the effects by contention at the network level 
on message exchange
latencies can in fact be very complex (or even non-feasible,
especially in virtualized cloud infrastructures), given the difficulty to gain access to
detailed information on the exact dynamics of messaging/network-level components \cite{CouceiroRR10}.

\remove{
In TAS, we exploit the availability of a complementary white-box model of a system’s
performance to formulate the machine-learning-based forecasting problem in a
way that differs significantly from traditional, pure black-box approaches. Conventional
machine-learning-based techniques (e.g., Shen et al. [2011]) try to forecast some
performance metric p2 in an unknown system configuration c2, given the performance
level p1 and the demand of resources d1 in the current configuration c1. In TAS, instead,
the analytical model can provide the machine learner with valuable estimates of the
demand of resources d2 in the target configuration c2. Specifically, we use the analytical
model to forecast what will be, in the target configuration c2, the rate of transactions
that will initiate a commit phase as well as the percentage of CPU time consumed by
the threads in charge of processing transactions.
As already mentioned, contention on the network layer can have a direct impact on
the latency of a key phase of the transaction’s execution, namely, the duration of the
commit phase, denoted as Tprep for 2PC and as Tcd for PB.
}

As already mentioned, contention on the network layer, and the associated message delivery delay, can have a direct impact on
the latency of two key transaction execution phases within the data grid, namely the
distributed commit phase, and the fetch of data whose copies are not locally kept by the cache server, given that the whole data-set might be only partially replicated across the nodes (e.g. for scalability purposes). These latencies, in their turn, may affect the rate of message exchange, and so the actual load on the messaging system (in the simulated configuration of the workload and for the specific data grid settings).

More in general, estimating (hence predicting) the message transfer delay while simulating some data grid system deployed over a specific networking software/hardware (virtualized) stack boils down in our approach to a non-linear regression problem, in which we want to learn the value of continuous functions defined on multivariate domains. Given the nature of the problem, we decided to rely
on the Cubist machine learning framework \cite{cubist}, which is a decision-tree regressor that
approximates non-linear multivariate functions by means of piece-wise linear approximations.
Analogously to classic decision-tree-based classifiers, such as C4.5 and ID3
\cite{c4.5}, Cubist builds decision trees choosing the branching attribute such
that the resulting split maximizes the normalized information gain. However, unlike
C4.5 and ID3, which contain elements in a finite discrete domain (i.e., the predicted
class) as leaves of the decision tree, Cubist places a multivariate linear model at each
leaf.

Clearly, the reliance on machine-learning requires building an initial knowledge base in relation to the networking dynamics of the target virtualized infrastructure, for which we need to simulate the behavior of some specific data grid system (or configuration) run on top of it.
This can be achieved by running (possibly once) a suite of
(synthetic) benchmarks that generate heterogeneous
workloads in terms of mean size of messages, memory footprint at each node, CPU utilization,
and network load (e.g. number of transactions that activate the commit phase per
second). As for this aspect, one could exploit some (open source) data grid system relying on the specific messaging layer for which the machine learner must provide the predictions. This approach looks perfectly suited for data-grid providers (namely for scenarios where the data-grid system is provided as a PaaS \cite{cloud-tm-site}), given that they can take advantage of (historical) profiling data related to specific (group) communication and messaging systems run on top of given (consolidated) virtualized platforms.

Also, it is well known that the selection of the features to be used by machine-learning
toolkits plays a role of paramount importance, since it has a dramatic impact on the
quality of the resulting prediction models. When performing such a choice, we took
two aspects into consideration: first, the set of parameters has to be large enough to
guarantee good accuracy, but at the same time it has to be small enough to
keep low the time taken by the machine learner to create its model and invoke it.

Second, the set of features has to be highly correlated to the parameters the machine
learner is going to predict, namely the message transfer delay across nodes within the system. In the following, we list the set
of features we selected, also motivating our choices:
\begin{itemize}
\item Used memory:  it has been shown that the memory footprint of
applications can affect significantly the performance of the messaging layer \cite{taas,CouceiroRR10}.
\item CPU utilization: this parameter is required given that the message delivery latency predicted by our machine learner
 includes a portion related to CPU processing (such as the marshalling/unmarshalling of the message payload).

\item The message size: this parameter is of course highly related to the time
needed to transmit messages over the (virtualized) networking infrastructure.
\item The number of message exchange requests per second: this parameter
 provides a good indicator of the network
utilization.
\end{itemize}

Clearly, predicting metrics such as the message delivery latency under a specific simulation scenario
depends on how the simulation model progresses, e.g., in terms of simulated system throughput and consequent
actual number of message exchange operations per second (see the last parameter listed above).
 These parameters, as well as others (like the average size of exchanged messages), are in their turn targeted in the estimation by simulation.
Hence they might be unknown at the time in which the machine
learner is queried during the simulation run.

\begin{figure}[t]
\centering
 \includegraphics[width=9cm]{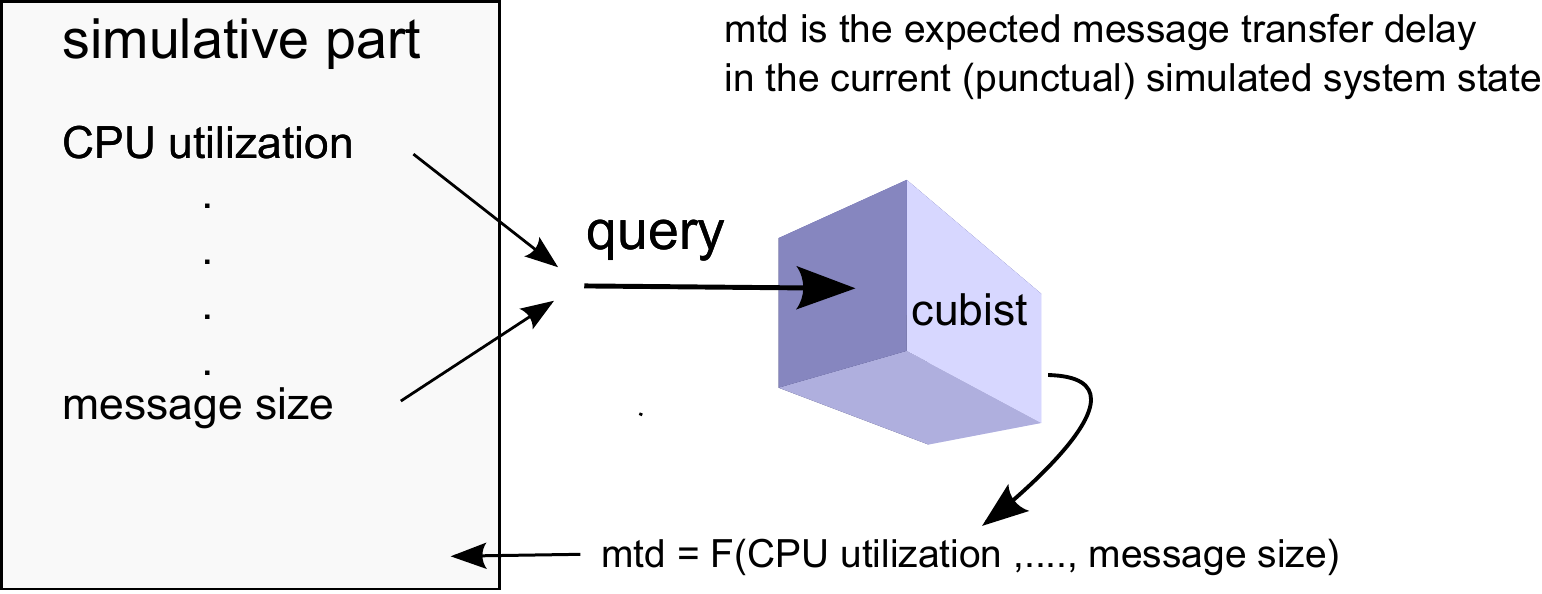}
\caption{Coupling of simulative and machine learning components.}
 \label{query}
\end{figure}

This problem is intrinsically solved by the specific way we couple simulative and machine learning components.
Particularly,
when a prediction on the delay of message delivery is required  for a specific message send operation, the simulative components
compute (estimate) the values
needed as input by the machine learning component, depending on the current simulated system state.
This is done easily and efficiently given that in our framework all the values of the parameters required in input by the machine learner (e.g. the current CPU utilization) to carry out its prediction are constantly updated, hence they are readily available. By using these values, the actual query to the machine learner is issued to determine the timestamp of the discrete-event associated with the message delivery along the simulation time axis. This coupling scheme is depicted in Figure \ref{query}, and the actual implementation
of this kind of interaction within our framework has been based on linking Cubist as a library directly accessible (invocable) by the simulation software.

This coupling approach leads the machine learner to output ``updated"
prediction for the message transfer delay (as a function of the message size), while the simulation run approaches the steady state value for the targeted parameters to be estimated (e.g. the system throughput, which may in turn depend on parameters like CPU usage). Hence, the process of ``rejuvenating"  the predictions by the  machine learner
ends upon converging towards the actual final estimation of the target parameters by the simulation run.

\section{Experimental Validation}
\label{data}



The skeleton operations described in the former session, such as the ones related to 2PC coordination, compose the foundational/base simulative model of our framework, (which users can extend and customize  to meet their needs).
For this reason, we have decided to validate the framework against real data achieved by running a data grid system exactly exploiting such an archetypal 2PC coordination paradigm.
  In particular, we present validation data obtained by comparing simulated performance results with the corresponding ones achieved by running the 2PC-based mainstream Infinispan data grid system by JBoss/Red-Hat \cite{infinispan-cache-mode}. Also, our experimentation has been based on a wide spectrum of system settings given that we consider large scale deployments on top of both public and private cloud systems. Finally, the workloads generated in our tests are based on various configurations of the YCSB benchmark by Yahoo \cite{ycsb}. Given that this benchmark has been devised just to assess (cloud suited) in-memory data stores, its employment further contributes to the relevance of the experimental configurations selected for validating the framework.


\remove{

 on top of a cloud based infrastructure hosted by Amazon EC2.
The used benchmark for  validation is a port of TPC-C \cite{tpcc} (already exploited in, e.g., \cite{icdcs2012}) that has been performed in order to allow this benchmark to be adapted to the $<key,value>$ data model (rather than the original relation model used in the benchmark specification).

}
\subsection{Overview of the Infinispan Data Grid Platform}

Infinispan is a popular open source in-memory data grid currently representing both the reference data platform and the clustering technology for JBoss, which is the mainstream open source J2EE application server.
Infinispan exposes a pure key-value data model (NoSQL), and maintains data entirely in main-memory relying on replication as its primary mechanism to ensure fault-tolerance and data durability. 
As other recent NoSQL platforms, Infinispan opts for weakening consistency in order to maximize performance. Specifically, it does not ensure serializability \cite{bernstein-book}, but only guarantees the Repeatable Read ANSI/ISO isolation level \cite{critique-ansi-sql}. At the same time, atomicity of distributed updates is achieved via 2PC. This is used to lock all the data object belonging to the write-set of the committing transaction, so as to atomically install the corresponding new data versions. The old committed version of any data object remains anyhow available for read operations until it gets superseded by the new one.

In the Infinispan version selected for our experiments, namely V5.1, the 2PC protocol operates according to a primary-owner scheme. Hence, during the prepare phase, lock acquisition is attempted at all the primary-owner cache servers keeping copies of the data objects to be updated.
If the lock acquisition phase is successful, the transaction originator broadcasts a commit message, in order to apply the modifications on these remote cache servers, which are propagated to the non-primary owners.

\subsection{Exploited Cloud Infrastructures}

The experimental test-bed for our validation study consists of a private and a public cloud infrastructure.
The Virtual Machines (VMs) deployed on both clouds are equipped with 1 Virtual CPU (VCPU) and 2GBs of RAM. They all run a Fedora 17 Linux distribution with kernel 3.3.4.

The private cloud consists of 140 VMs
deployed over a cluster composed of 18 machines
equipped with two 2.13 GHz Quad-Core Intel(R) Xeon(R)
processors and 32 GB of RAM and interconnected via a
private Gigabit Ethernet. 
Openstack Folsom is employed to regulate the provisioning of resources and Xen is used as virtualization software.
The public cloud consists of 100 VMs,
deployed over the FutureGrid India infrastructure \cite{futuregrid},
which exploits the Openstack Havana virtualization software.

\subsection{Workload Configurations}

We rely on three different workload configurations provided by YCSB, which we refer to as A, B and F.
Workload A has a mix of 50\% read and 50\% write (namely update) transactions; workload
B contains a mix with 90\% read and 10\% write transactions, while in workload F records
are first read and then modified within a transaction.
Also, we have ran experiments with two different data access profiles. In the first case, 
the popularity of data items follows a
zipfian distribution with YCSB's
zipfian constant set to the value 0.7.
In the second one, which we name hot spot case, 99\% of the data requests are issued against
the 1\% of the whole data set. A total amount of 100000 data objects constitutes the data set in all the experiments.

In the plots, we will refer to a specific workload configuration
using the notation N-D-P-I, where: `N' refers to the original
workload's YCSB notation \cite{ycsb}; `D' is the number of distinct
data items that are read by a read-only transaction; for update
transactions, it is the number of distinct data items that are
written (for the `F' workload, which exhibits a read-modify
pattern of update transactions, any accessed data is both read and written); 'P' encodes the data access pattern (`Z' stands for zipfian, `H' for hot-spot);
finally, `I' specifies the cloud infrastructure over which the
benchmark has been run (`PC' stands for private cloud, `FG' for
FutureGrid).

\subsubsection{Achieved Results}

All the above illustrated workload configurations have been run on top of the selected cloud systems while scaling the number of VMs, and relying on a classical consistent hashing \cite{Karger97consistenthashing} scheme for placing the data copies across the servers.
The run outcomes  have been exploited both to collect statistically relevant values for core performance parameters in the real system deploy and to determine the value of the parameters input parameters for the simulated data grid.
%
%
Specifically, we instrumented the YCSB implementation, as well as the Infinispan data grid system in order to be able to measure (for the different workload configurations and system deployments) a wide set of parameters, the most relevant of which 
are listed 
  in Table \ref{parameters}. The latter all refer to CPU demand for the different modeled activities at the cache servers, given that networking/messaging (expected) delays across the servers do not represent input parameters to the discrete event models, and  are instead predicted by the Cubist machine learning component while the simulation run is in progress. To this end, the knowledge base to be acquired by Cubist consists of simple textual files, which have been populated while profiling netwoking/messaging dynamics in real runs of the system.

We note that the parameters reported in Table \ref{parameters}, together with others, such as the data placement across the different cache servers (namely the association of replicated $\langle key,value\rangle$ pairs to the cache servers) and the inter-time between subsequent {\sf put}/{\sf {get}} commands by the client, can be the object of tuning by the user, e.g., for what-if analysis purposes. Given that in this study we have a different target, namely the validation of the framework taking
the selected set of workload/deploy configurations as the reference, we fixed the tunable parameters' values to the ones measured/set for the corresponding target configuration used as an individual validation sample.

As a final preliminary note, in the real system the workload generator has been deployed as  a thread running on each VM, which injects requests against the collocated Infinispan cache sever instance, in closed loop. Consequently, in the simulation model configuration, no networking/messaging delays have been modeled between clients and cache server instances. Yet, the (simulated) networking/messaging system plays a core role in the data exchange and coordination across the  different cache server instances. This well fits the relevant scenarios where the focus of performance analysis/prediction is on sever side infrastructures.

\begin{table}
\caption{Measured Parameter Values (to Configure the Simulation Models).}
\label{parameters}
\centerline{
\begin{tabular}{|l|}
\hline
\hline
{\tt local\_tx\_get\_cpu\_service\_demand}  \\
{\tt local\_tx\_put\_cpu\_service\_demand}  \\
{\tt local\_tx\_get\_from\_remote\_cpu\_service\_demand}   \\ 
{\tt tx\_send\_remote\_tx\_get\_cpu\_service\_demand}    \\
{\tt tx\_begin\_cpu\_service\_demand}   \\
{\tt tx\_abort\_cpu\_service\_demand}   \\
{\tt tx\_prepare\_cpu\_service\_demand}   \\
{\tt distributed\_final\_tx\_commit\_cpu\_service\_demand}  \\
\hline
\end{tabular}
}
\end{table}


\begin{figure*}
\centering
 \subfigure[Throughput]
   {\includegraphics[width=0.5\textwidth]{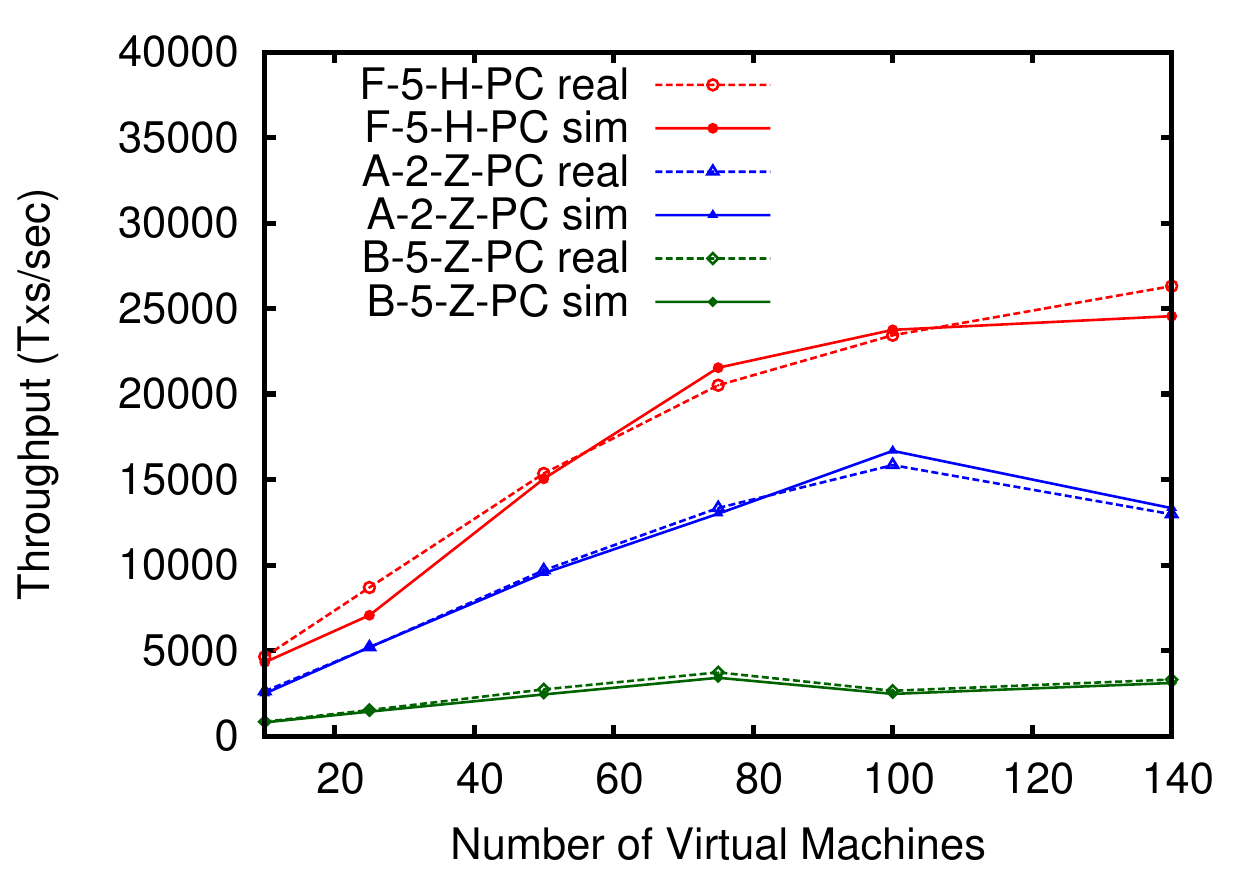}}%
 \subfigure[Commit Probability]
   {\includegraphics[width=0.5\textwidth]{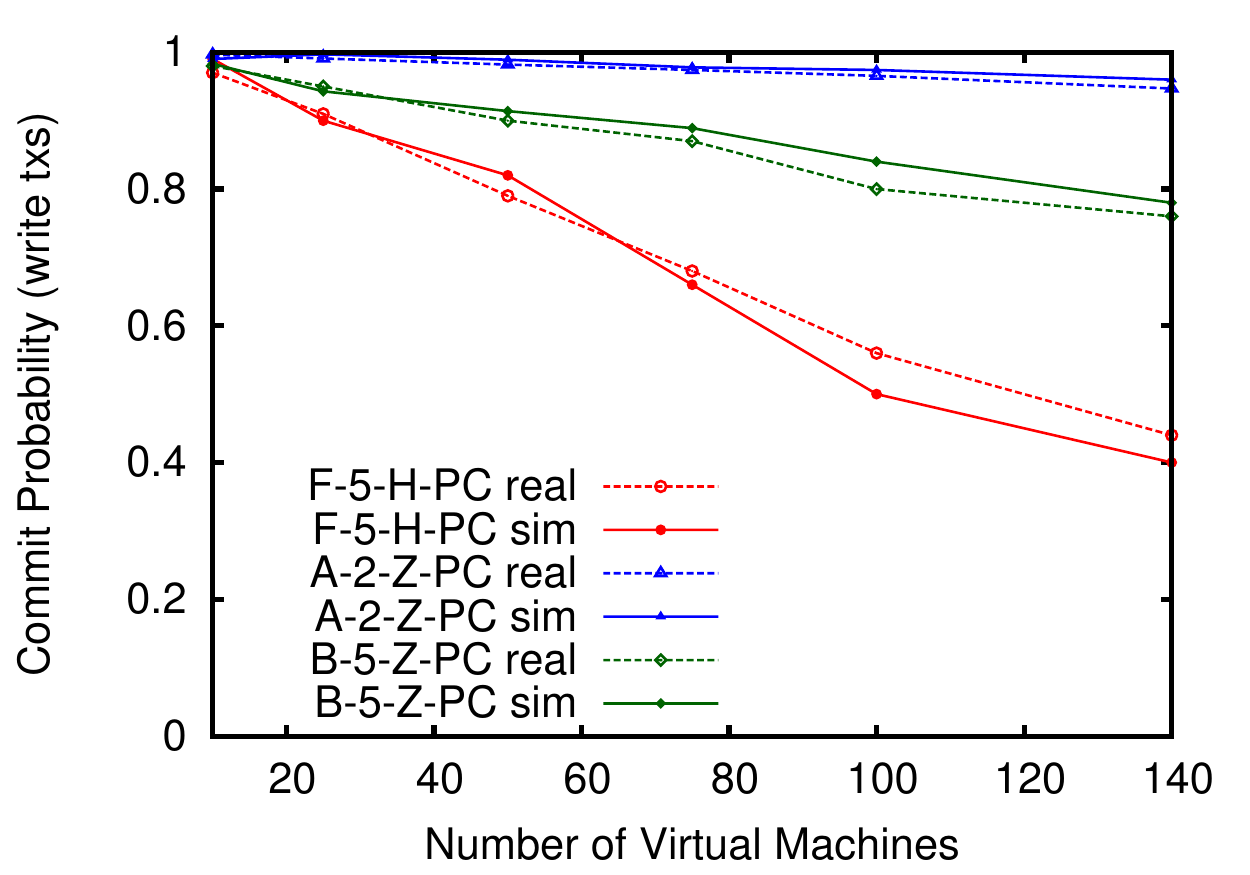}}
\subfigure[Response Time (Read Only)]
   {\includegraphics[width=0.5\textwidth]{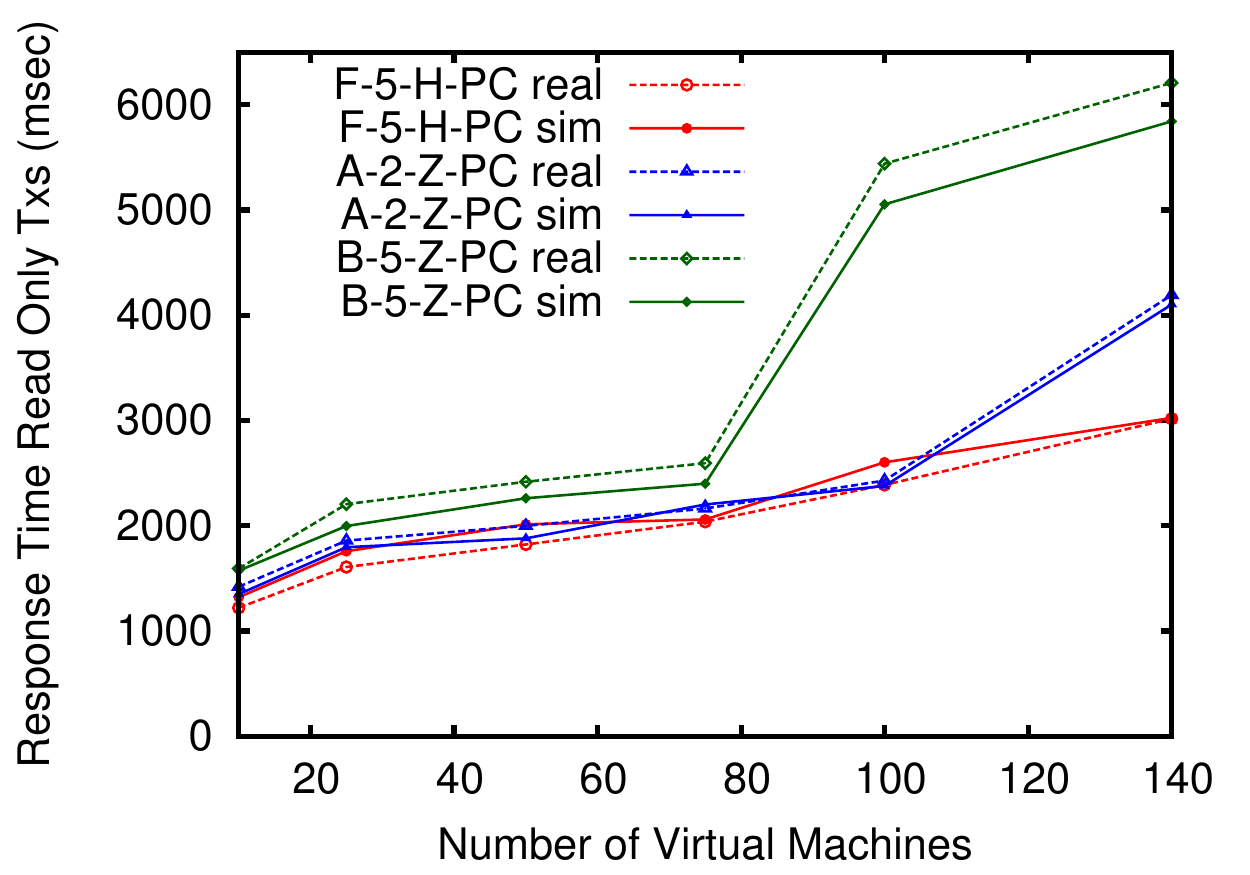}}%
 \subfigure[Response Time (Read/Write)]
   {\includegraphics[width=0.5\textwidth]{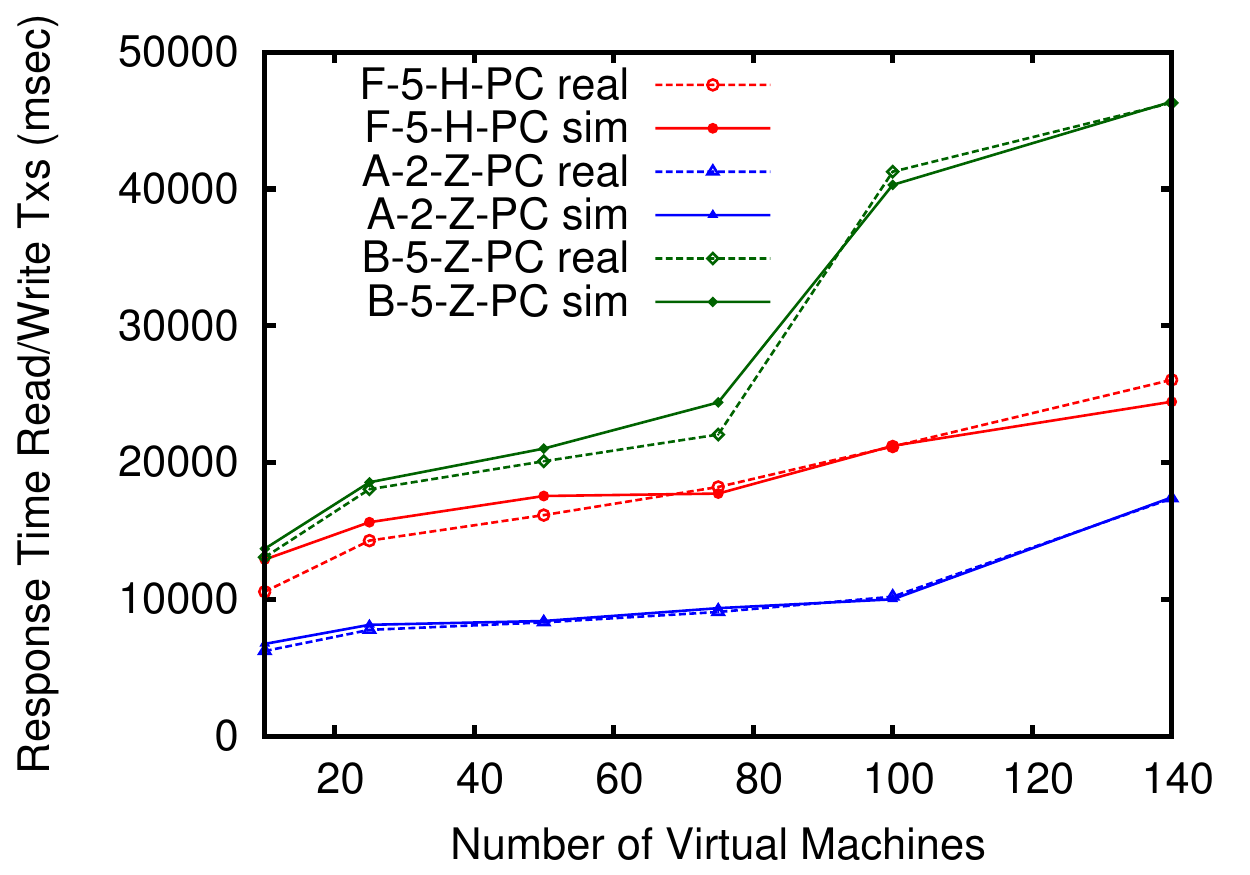}}
 \caption{Results for Deploy on the Private Cloud (up to 140 VMs).}
\label{pc}
\end{figure*}

\begin{figure*}
\centering
 \subfigure[Throughput]
   {\includegraphics[width=0.5\textwidth]{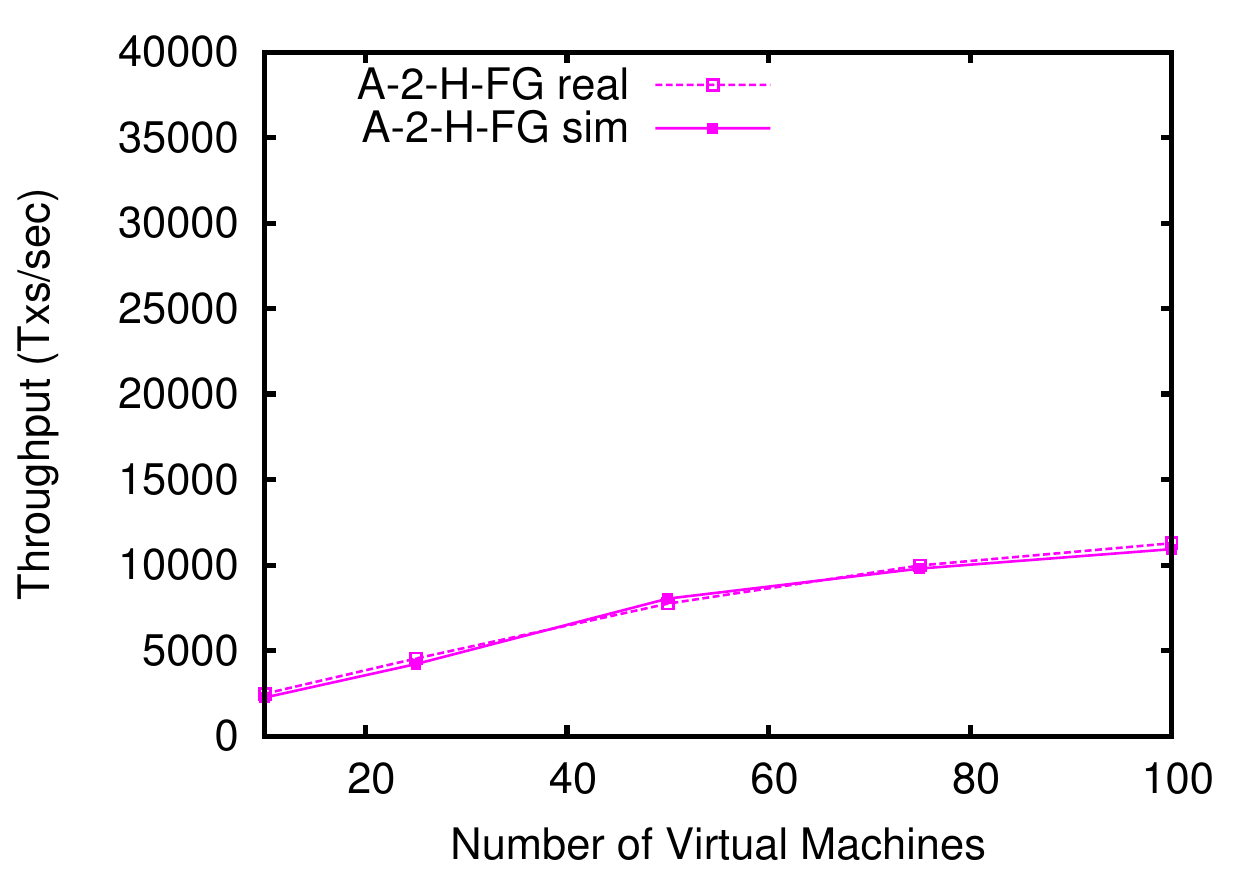}}%
 \subfigure[Commit Probability]
   {\includegraphics[width=0.5\textwidth]{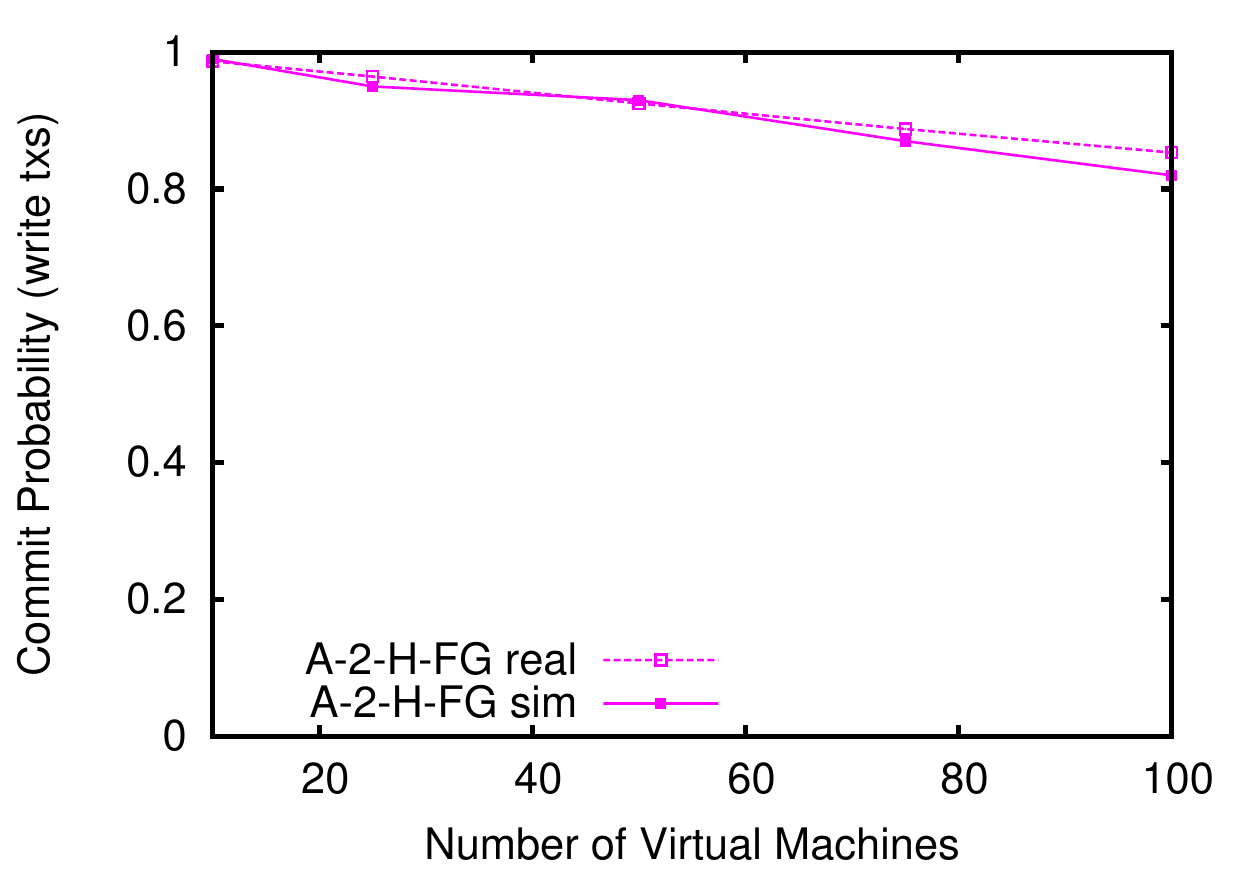}}
\subfigure[Response Time (Read Only)]
   {\includegraphics[width=0.5\textwidth]{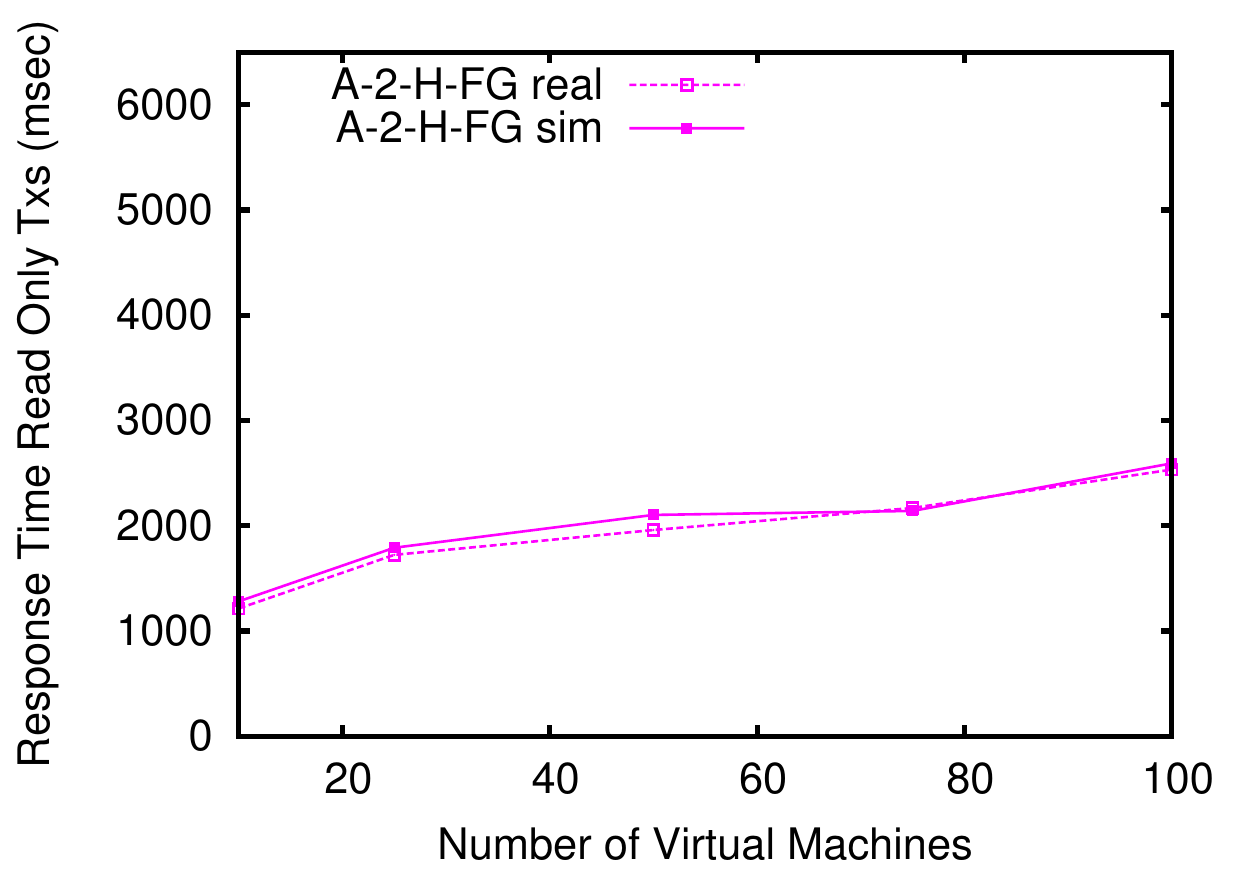}}%
 \subfigure[Response Time (Read/Write)]
   {\includegraphics[width=0.5\textwidth]{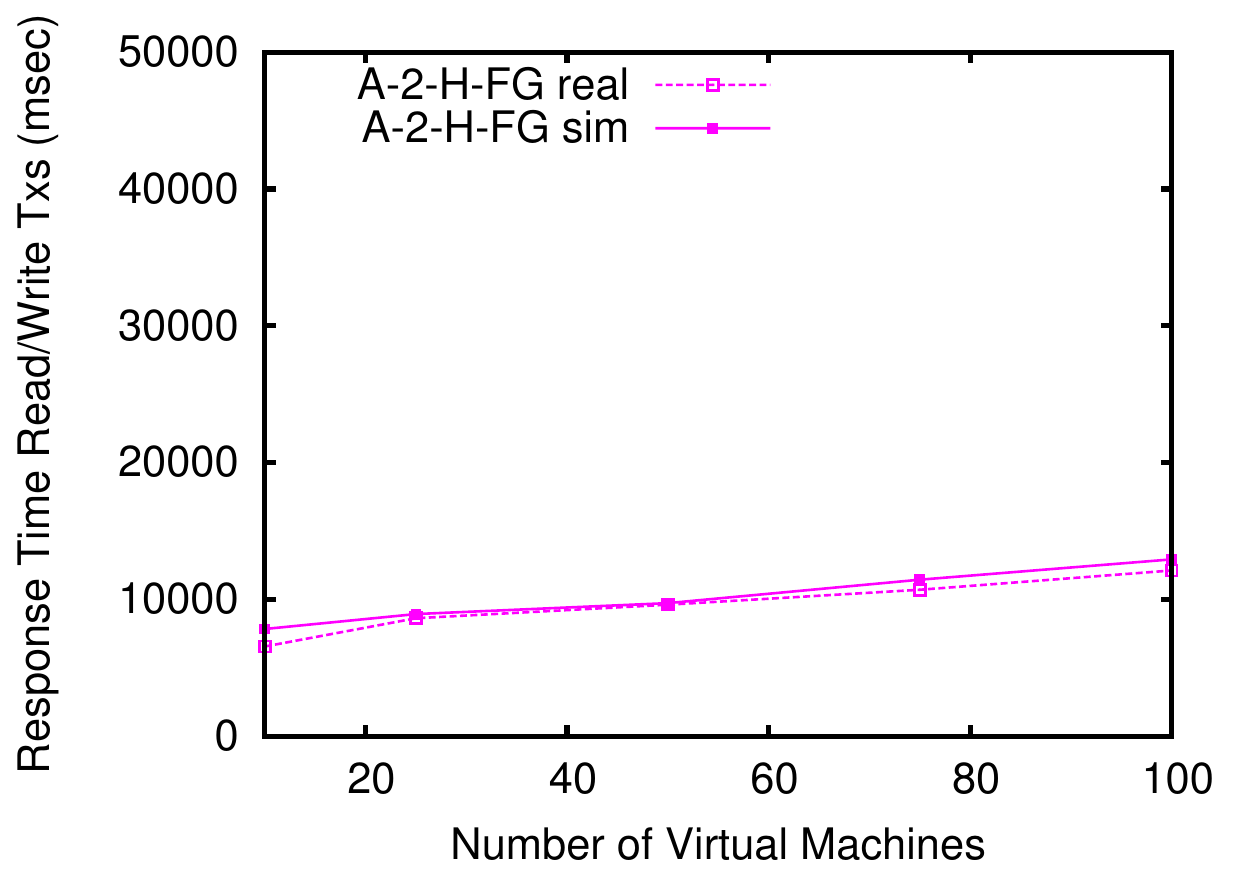}}
 \caption{Results for Deploy on FutureGrid (up to 100 VMs).}
\label{fg}
\end{figure*}

The validation has been based on measuring the following set of Key Performance Indicators (KPIs), and comparing them with the ones predicted via simulation: (i) the system {\em throughput}, (ii) the transaction {\em commit probability} (this parameter plays a role for update transactions, given that read-only transactions are never aborted by the concurrency control algorithm considered in this study), and (iii) the {\em execution latency} of both read-only and update transactions. The first KPI provides indications on the overall behavior of the system, and hence on how accurate is the corresponding prediction by the framework-supported models. The second one is more focused to the internal dynamics of the data grid system (e.g. in terms of the effects of the distributed concurrency control mechanism), which have anyhow a clear effect on the final delivered performance. Finally, the execution latency of the different types of transactions has been included in order to provide indications on how the simulative models are able to reliably capture the dynamics of different kinds of tasks (exhibiting different execution patters) within the system. In fact, read-only transactions can require remote data fetches across the cache severs but, differently from update transactions, they entail no 2PC step.

The results for the case of data grid deploy on top of the private cloud system are reported in Figure \ref{pc}. For all these tests we considered a configuration where each data-object is replicated two times across the cache servers, which is a typical settings allowing for system scalability, especially in contexts where genuine distributed replication protocols are used to manage the data access \cite{AguileraMSVK09,DBLP:conf/middleware/PelusoRQ12} (\footnote{A distributed transactional replication protocol is said to be genuine if it requires contacting only the nodes handling the data copies accessed by a transaction in order to manage any phase of the transaction execution, including its commit phase.}). Therefore, this value well matches the nature of this particular validation study, given that we consider deployments on large scale infrastructures (up to 140 VMs). Also, the variance of simulation results across different runs (executed with different random seeds) is not explicitly plotted given that the obtained simulative values were quite stable, differing by at most 10\%.

By the plotted curve we can see how the KPIs' values predicted via simulation have a very good match with the corresponding ones measured in the real system, at any system scale. As an example, the maximum error on the overall throughput prediction is bounded
by 20\%, as observed for the configuration F-5-H-PC when running on top of 25 VMs. However, except for such a peak value, the error in the final throughput prediction is in most of the cases lower than 5\%. Similar considerations can be drawn for the other reported KPIs.

Another interesting point is related to the fact that the simulative models are able to correctly capture the real system dynamics when changing the workload. As an example, while we observe higher commit probability for an individual run of an update transaction in the scenario with the 50\%/50\% read/write mix and zipfian data accesses, the hot spot configuration allows for higher throughput values even though the update transaction commit probability is lower. This is clearly due to the fact that in the used hot spot configuration only 5\% of the whole workload consists of update transactions that, although being subject to retries due to aborts with non-minimal likelihood (especially at larger system scales), impact the system throughput in a relatively reduced manner.

The results achieved for the case of deploys on top of the FutureGrid public cloud systems, which are reported in Figure \ref{fg}, additionally confirm the accuracy of the models developed via the framework. In these experiments we further enlarge the spectrum of tested scenarios not only because we move to a public cloud, but also because (compared to the case of private cloud deploy)
we consider a different value for the replication degree of data-objects across the servers, namely 3.
This value leads to the scenario where fault resiliency is improved over the classical case of replication on only 2 cache servers, which is the usual configuration that has been considered in the previous experiments. By the data we again observe very good match between real and simulative results. Further, similarly to the previously tested configurations, such a matching is maintained at any system scale, and, importantly, when the actual dynamics of the data grid system significantly change while scaling the system size. In fact, we observe that the commit probability of update transactions significantly changes when increasing the system scale. This phenomenon, and its effects on the delivered performance, are faithfully captured by the simulator. This is a relevant achievement when considering that the workload used for the experiments on top of the FutureGrid public cloud system has been based on a 50\%/50\% read/write transactions mix, which leads transaction retries to play a relevant role on the final performance given that half of the workload can be subject to abort events, which become increasingly frequent at larger scales of the system.

\section{Conclusions}

Optimized exploitation of cloud resources is a core topic to cope with, in any scenario. In this article we have presented a simulation framework for predicting the performance of cloud in-memory data grid systems which can be used for, e.g., what-if analysis aimed at the identification of the configurations (such as the number of virtual machines to be employed for hosting the data grid system under a specific workload profile) optimizing specific cost-vs-benefit tradeoffs.
 The design of the discrete event simulative framework has been based on the use of flexible skeleton models, which can be easily extended/specialized to capture the dynamics of data grid systems supporting, e.g., different distributed coordination schemes across the cache servers in order to guarantee specific levels of consistency in the transactional manipulation of data. The adequacy of the framework, and of its model instances, in predicting the dynamics of data grid systems hosted in cloud environments is a result of the combination of the discrete event simulative approach with machine learning. In our framework architecture, the latter modeling technique is used to predict the dynamics at the level of networking/messaging sub-subsystems which, in cloud contexts, are typically unknown in terms of their internal structure and functioning, and are therefore difficult to be reliably modeled via white-box approaches.

We have also presented a validation study where the simulation output by the framework has been compared with real data related to the execution of a mainstream open source data grid system, namely Infinispan by JBoss/Red-Hat, deployed on both a private and a public cloud infrastructure. This validation study of the simulative model of Infinispan has been based on large scale deploys on top of up to 140 Virtual Machines, and using the YCSB benchmark by Yahoo, in different configurations, as the generator of the test-cases workload profiles. By the data, the accuracy of the simulations in estimating core parameters such as the system throughput has been on the order of at least 80\%, and on the order of 95\% on the average, for all the tested configurations. Finally, the framework has been released as an open source package available to the community.


\end{document}